\documentclass[crcready]{jfm}

\usepackage{graphicx}
\usepackage{comment}
\usepackage{newtxtext}
\usepackage{newtxmath}
\usepackage{natbib}
\usepackage{hyperref}
\hypersetup{
    colorlinks = true,
    urlcolor   = blue,
    citecolor  = black,
}

\newcommand{\RomanNumeralCaps}[1]
\linenumbers

\title{On internal wave whispering gallery modes in channels and critical-slope wave attractors}

\author{N. Bratspiess\aff{1}
  \corresp{\email{bratspiess@mail.tau.ac.il}},
  L.R.M. Maas\aff{2}
 \and E. Heifetz\aff{3}}

\affiliation{\aff{1}Department of Physics and Astronomy, Tel Aviv University
\aff{2}Institute for Marine and Atmospheric Research Utrecht, Utrecht University
\aff{3}Department of Geophysics, Tel Aviv University}

\begin{document}
\maketitle

\begin{abstract}
Internal waves are an important feature of stratified fluids, both in oceanic and lake basins and in other settings. Many works have been published on the generic feature of internal wave trapping onto planar wave attractors and super-attractors in 2\&3D and the exceptional class of standing global internal wave modes. However, most of these works did not deal with waves that escape trapping. By using continuous symmetries we analytically prove the existence of internal wave Whispering Gallery Modes (WGMs), internal waves that propagate continuously without getting trapped by attractors. WGMs neutral stability with respect to different perturbations enable whispering gallery beams, a continuum of rays propagating together coherently. The systems' continuous symmetries also enable projection onto 2D planes that yield effective 2D billiards preserving the original dynamics.

By examining rays deviating from these WGMs in parabolic channels we discover a new type of wave attractor which is located along the channel instead of across it as in previous works. This new wave attractor leads to a re-understanding of WGMs as sitting at the border between the two basins of attraction. 
 
Finally, both critical-slope wave attractors and whispering gallery beams are used to propose explanations for along-channel energy fluxes in submarine canyons and tidal energy intensification near critical slopes. 
 
\end{abstract}

\begin{keywords}
Authors should not enter keywords on the manuscript, as these must be chosen by the author during the online submission process and will then be added during the typesetting process (see \href{https://www.cambridge.org/core/journals/journal-of-fluid-mechanics/information/list-of-keywords}{Keyword PDF} for the full list).  Other classifications will be added at the same time.
\end{keywords}

{\bf MSC Codes }  {\it(Optional)} Please enter your MSC Codes here

\section{Introduction}
\label{sec:headings}

The interiors of density-stratified or rotating fluids present anisotropic media \citep{Maas2022}. These allow for stable hydrostatic and cyclostrophic equilibria respectively. In the former equilibrium, the fluid density increases monotonically in the direction of gravity. In the latter equilibrium, the angular momentum increases radially outwards from the rotation axis. 
The single direction into which gravity or rotation axis points (vertical, say) is distinct from the two horizontal orthogonal directions. In a field of gravity, it is energetically costly to displace fluid parcels vertically. In rotating fluids, this applies to radial displacements. The disturbance is restored by gravity or centrifugal/centripetal forces, as well as by a pressure gradient force, imposed by neighboring fluid. 

In the fluid interior, anisotropy manifests itself in disturbances that propagate as internal waves under a fixed angle relative to gravity or rotation axis \citep{gortler1943schwingungserscheinung,oser1958experimentelle}. The wave’s frequency, restricted by an upper threshold imposed by stratification and/or rotation \citep{gerkema2008introduction}, determines this inclination. Upon reflection from a solid boundary, these waves preserve their frequency and hence their inclination, thus obeying a non-specular type of reflection in the vertical \citep{bretherton1964low}. This implies that in an enclosed fluid domain, possessing sloping boundaries, reflecting waves exhibit wave focusing and defocusing. In a two-dimensional enclosure, focusing dominates over defocusing. Generically (i.e. for nearly any wave frequency below the threshold) internal waves will approach a limit cycle \citep{bretherton1964low, stewartson1971trapped,stewartson1972trapped} – a wave attractor \citep{Maas_Lam_1995}. In ideal fluids, this attractor represents a spatial singularity, where gradients in pressure increase without bound. This evokes a nonlinear adjustment in which a new pair of waves is generated via a triadic resonant instability \citep{dauxois2017energy}. In real, viscous fluids, this process is accompanied or superseded by the development of free (internal) viscous boundary layers surrounding the wave attractor \citep{RIEUTORD_VALDETTARO_2010,hazewinkel2008observations,Le_Dizès_2015,he2022internal}. 

When the shape of the two-dimensional fluid domain has residual symmetries (such as, e.g., a trapezoidal basin), for certain distinct wave frequencies regular wave modes exist \citep{Maas_Lam_1995}. In such ‘globally resonant modes’, wave focusing is exactly balanced by subsequent defocusing. These modes resemble the classical, complete orthogonal set of eigenmodes encountered in elliptic problems where the spatial structure is governed by a Helmholtz equation \citep{Riley_Hobson_Bence_2006}. It is therefore tempting to search for such ‘eigenmodes’ in density-stratified and rotating three-dimensional fluids, also, regardless of the shape of the fluid domain. This search is fueled by the discovery of such eigenmodes in rotating spheres and then ellipsoids \citep{bryan1889vi,ZHANG_EARNSHAW_LIAO_BUSSE_2001,de2025spectrum}. Due to the symmetry of the sphere, these exist because the governing equation is solvable by separation-of-variables. But, this property is in general lost in the astrophysically and geophysically relevant spherical shell, obtained by adding an inner sphere to the fluid domain. Wave attractors still prevail. An exception is formed by the spin-over modes \citep{hough1895xii,LEBARS201048}, consisting in fluid displacements that strictly retain their radial position. Never meeting the concentric inner and outer spherical boundaries, these avoid (de)focusing reflections. 

Still, could there exist wave modes in the three-dimensional fluid domain that avoid trapping onto wave attractors? Due to limitations of the separation-of-variables method employed to construct globally resonant wave solutions, recourse is here taken to exploring short-wave ray dynamics, comparable to a geometric optics approach. This assumes that internal waves are short compared to basin scales. In the fluid interior, these waves follow straight lines along the energy propagation direction imposed by the group velocity. Upon meeting a boundary, obliquely incident waves reflect and instantaneously refract \citep{Phillips1963}, i.e. they change their horizontal propagation direction. This is because in an ideal fluid, the combined fluid particle motion along incident and reflected rays needs to vanish in a direction normal to the boundary. Hence, decomposing the fluid motions into components in the vertical plane normal to the boundary and a component parallel to the boundary, the latter is unchanged during the reflection process while the former is subject to (de)focusing reflections. Upon iteration of this process, the internal wave ray bounces around the fluid domain. Many rays then reach a limit cycle, a wave attractor. In three-dimensional domains, because of basin symmetries (such as in a paraboloid), this can be a two-dimensional attracting manifold \citep{maas1997observation, maas2005wave}. But when this symmetry is absent, the waves can be further focused onto a one-dimensional manifold, a ‘super-attractor’ \citep{pillet2019internal, favier2024inertial}.

Interestingly, there exist rays that upon iteration avoid trapping and that resemble whispering gallery modes (WGMs), initially also termed ‘edge waves’ \citep{manders2004three,drijfhout2007impact,rabitti2013meridional, rabitti2014inertial}. These modes were first discovered in acoustics \citep{Rayleigh1878Theory}, where they represent sound waves that are guided along an outer wall. These spread much less than sound emitted into the interior of an acoustic chamber due to the lack of geometric spreading when propagating on a surface of lower dimension (the gallery's wall counter to its interior). WGMs are also found in other media, such as electromagnetic waves \citep{mie1908beitrage,Debye1909WGM}, matter waves for neutrons \citep{nesvizhevsky2010neutron}, and more. For internal waves, WGMs are instead guided by the critical line, the line on-which boundary slope equals to the internal wave ray inclination. Although they share a name the mechanism enabling attenuation-less propagation in internal wave WGMs differs from the one in acoustic ones. Here we study WGMs in several basin shapes. This is done by geometric construction of periodic orbits and by examining deviations from said periodic orbits. By perturbing these orbits we find cases that represent isolated periodic orbits, as well as cases having some meta-stability, that might be used to collectively construct a spatially extended beam of internal waves. Moreover, these deviations lead us to discover new kinds of attractors not yet predicted, which lay parallel to the channel's direction instead of perpendicular to it. We use these new attractors to explain energy accumulation near critical slopes in submarine settings. In Appendix A. we deal with the case of rotationally symmetric basins.  
\section{3D internal wave ray reflection law}
\begin{figure}
  \centerline{\includegraphics[width=12cm]{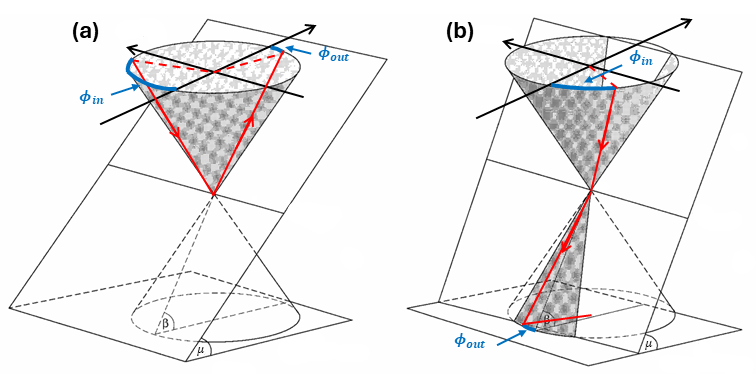}}% Images in 100% size
  \caption{ (a) Sub-critical internal wave ray reflection. (b) Super-critical internal wave ray reflection. Red arrows indicate wave energy propagation direction. Adapted from \cite{thorpe1997interactions}. Angles are presented with respect to both up-slope and down-slope although measured with respect to up-slope exclusively.}
\label{fig:1}
\end{figure}

Our system consists of a 3D inviscid Boussinesq fluid of linear stratification and uniform rotation, with a squared Brunt-Vaisala frequency $N^2 = -\frac{g}{\rho_0} \frac{d \rho(z)}{dz}$ and  Coriolis parameter $f$. Here, $g$ denotes the accelaration of gravity, while $\rho_0$ and $\rho$ denote the spatio-temporally uniform and static $z$ dependent parts of the density field respectively. We assume that all fields are periodic in time and scale as $e^{-i\omega t}$. After some manipulation, using subscript derivatives, we arrive at the linearized Poincar\'e equation for the spatial pressure perturbation \citep{maas2005wave}.
\begin{equation}
    p_{xx}+p_{yy}-\gamma^2(\omega)p_{zz}. 0\end{equation}
Its corresponding dispersion relation reads
 \begin{equation}
 \gamma^2(\omega) =
    \frac{\omega^2-f^2}{N^2-\omega^2}=\frac{k_x^2+k_y^2}{k_z^2}=\tan^2{\beta}
\end{equation}
in terms of wave vector 
\begin{equation}
    \boldsymbol{k} = (k_x,k_y,k_z) \propto (\cos{\phi_k},\sin{\phi_k},\pm\gamma^{-1})
\end{equation}
and where $\beta$ denotes the wave vector's angle from the vertical. Ray inclination $\gamma$ depends on the frequency of the wave, $\omega$. By scaling horizontal coordinates $x'=x/L,y'=y/L$ by horizontal domain scale $L$ and by stretching the vertical coordinate, $z'=\gamma z/L$, rays of any frequency $f<\omega<N$, or vice versa $f>\omega>N$, follow characteristics with inclination 1. When examining internal wave rays in a closed basin this scaling and stretching comes at the price of changing  its characteristic height $H$ to the stretched height $\tau=\gamma H/L$. A full derivation of equations of motion and stretched height from the Boussinesq approximation can be found in \citep{Maas_Lam_1995}. $\phi_k$ is the horizontal angle of the wave, relative to the positive $x$ direction, and the $\pm$ sign is decided by whether the wave vector points up or down. The vertical directions of group velocity and phase velocity are orthogonal to each other. When $N>f$, a wave with a wave vector component that points upward will propagate energy with a downward component and vice versa. For $N<f$, phase and energy propagate in the same vertical direction but opposite horizontal direction. When hitting a locally linear wall with slope $s = \tan{\mu}$, a ray characterized by wave vector $\boldsymbol{k}$ reflects to a different direction $\boldsymbol{m}$. In terms of a scaled wall slope
    \begin{equation}
        \eta\equiv\frac{s}{\gamma}
    \end{equation} the reflection law determining the horizontal exit angle $\phi_{out}$ can be presented in 3 different, equivalent ways.
    \begin{subeqnarray}
   \tan{\phi_{out}} &=& \frac{(1 - \eta^2)\sin{\phi_{in}}}{(1 + \eta^2)\cos{\phi_{in}} + 2sign(k_z)\eta}\\
  \sin{\phi_{out}}&=&sign(1-\eta)\frac{(1-\eta^2)\sin{\phi_{in}}}{(1 + \eta^2) + 2sign(k_z)\eta\cos{\phi_{in}}}\\
   \cos{\phi_{out}}&=&sign(1-\eta)\frac{(1+\eta^2)\cos{\phi_{in}} + 2sign(k_z)\eta}{2sign(k_z)\eta\cos{\phi_{in}}  +(1+\eta^2)}
\end{subeqnarray} 
    The bottom slope is called subcritical when the bottom slope is less than the ray slope, $\eta<1$ and supercritical if it is steeper, $\eta>1$. This is illustrated in panels (a) and (b) of Fig.~\ref{fig:1}, where $\phi_k = \phi_{in}$ and $\phi_m = \phi_{out}$ represent the horizontal angles, relative to the up-slope $x$-direction, of the rays that are incident and refracted/reflected from a sloping wall, respectively. Note the reciprocity of these angles when reversing their propagation direction, indicated by arrows. 
    
\section{Internal wave rays in parabolic channels}
In this paper we look at internal wave rays in infinite channels. The channels we are interested in are symmetric along the along-channel direction, and have  a piecewise continuous transverse bottom profile and a flat surface at which we employ the rigid lid approximation. In these channels of stretched depth $\tau$, we examine rays of inclination $\gamma=1$. From here on, without loss of generality, all channels lay along the $\hat{y}$ direction.

As the system is symmetric along the $y$-axis, it is natural to look at the two-dimensional projection on the $x-z$ plane (see Fig.2). In that plane, the ray takes an effective inclination 
\begin{equation}
    \gamma'\equiv\frac{-1}{\cos{\varphi}},
\end{equation}
where $\phi=\varphi$ for a given ray propagating through the parabolic channel. 
We note that a ray propagating parallel to the $y$-direction (i.e. $\varphi=\pm \pi/2$) has an effective inclination $\gamma'=\infty$, and a ray propagating parallel to the $x$-direction (i.e. perpendicular to the sloping bottom, $\varphi=0 \mbox{ or } \pi$) has $\gamma'=\pm1$.

In a sense, this projection transforms the three-dimensional internal wave ray billiard system to a two-dimensional billiard with a complex reflection law. A periodic orbit of the 2-D billiard in this $x-z$ projection plane, where rays constituting the periodic orbit have values of $\gamma'\ne 1$, proves the existence of an internal wave ray in three dimensions that forever propagates along the channel without getting trapped onto a wave attractor: an internal wave whispering gallery mode (WGM); see Fig.~\ref{fig:2}.

\subsection{WGM existence}
When examining cross sections of a parabolic channel, it is natural to use the parametrization used in \citep{Maas_Lam_1995}. The parabolic cross section is characterized by two parameters, the half-width of the channel, with a scaling equal to 1, and its stretched depth $\tau$.  As in parabolic channels with $\tau<0.5$ all reflections are subcritical, meaning that all rays get attracted to the beaches, we focus on parabolic channels with $\tau>0.5$. This also ensures the existence of a critical slope in the channel: the line where the local bottom slope equals the ray slope.

\begin{figure}
  \centerline{\includegraphics[width=13.6cm]{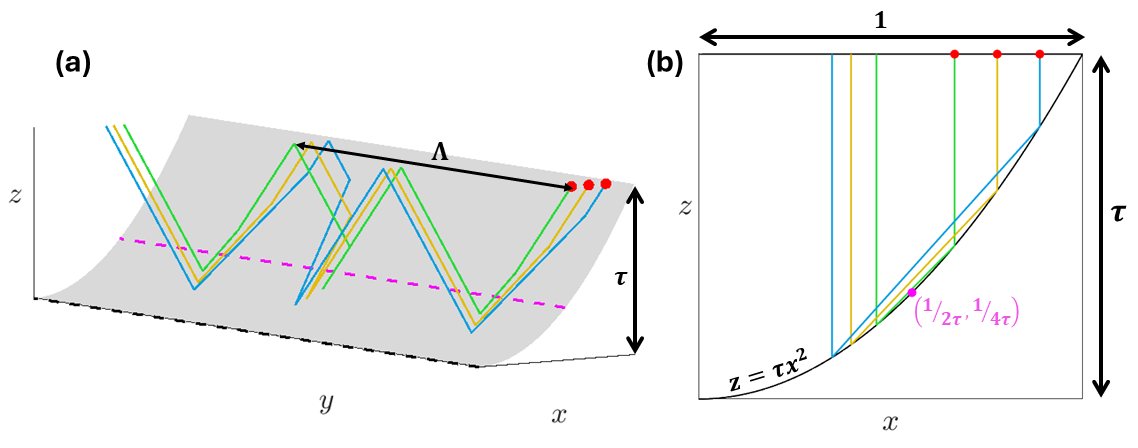}}% Images in 100% size
  \caption{(a) WGMs in the right half of a parabolic channel launched from the red dots parallel to the $y$-direction. The critical line is shown in magenta. (b) Two-dimensional projection of (a) onto the $x-z$ plane. vertical lines correspond to rays propagating parallel to the channel.}
\label{fig:2}
\end{figure}

Although different WGMs may also be constructed, inspired by the ones proposed by \citep{drijfhout2007impact} we examine rays launched down-channel from the rigid lid surface, i.e. parallel to the $y$-direction. A ray launched from $x_+>x_{crit}=\frac{1}{2\tau}$ hits a wall with slope $s=2\tau x_+$, which focuses it to the angle defined by 
\begin{equation}
    \cos{\varphi}=\frac{-2s}{1+s^2}=\frac{-4\tau x_+}{1+(2\tau x_+)^2}\Longleftrightarrow x_+=x_{crit}(\frac{-\cos{\varphi}}{1+\sin{\varphi}})
\end{equation}
After propagating in the 2D projection billiard with effective inclination $\gamma'\equiv\frac{-1}{\cos{\varphi}}$, the ray intersects the channel at 
\begin{equation}
    x_-=x_{crit}(\frac{-\cos{\varphi}}{1-\sin{\varphi}})=\frac{x_{crit}}{s}.
\end{equation}
This makes sense as $s>1$. At that point, the slope $2\tau x_-$ reflects the ray back to being parallel to the channel due to the reflection law being symmetric, a ray backtracking will perfectly retrace its path. After the ray hits $(x_-,\tau x_-^2)$ and a subsequent surface reflection, it reverses its own path back to $x_+$ and parallel to the $y$-direction. Therefor we have found a continuum of periodic orbits in the 2D billiard, meaning we have found a continuum of WGMs in parabolic channels. These WGMs can be seen in Fig 2, both in the 3D system and in its 2D projection. 

These WGM pairs of
\begin{equation}
    x_{\pm}=x_{crit}(\frac{-\cos{\varphi}}{1\pm\sin{\varphi}})
\end{equation}
are capped by the width of the channel $x_+<1$ that in turn dictates $x_->\frac{1}{4\tau^2}$. Rays launched outside this bound get trapped onto wave-attractors.

We define the WGM step size as the distance traveled down the channel during a single period in the 2D projection billiard. As can be seen in Fig.~\ref{fig:2}, all such WGMs share a step size:
\begin{equation}
    \Lambda=4\tau(1 - \frac{1}{4\tau^{2}}).
\end{equation}
\subsection{Stability analysis}
\subsubsection{Spatial perturbations}
As we have established the existence of launching conditions leading to periodic orbits, we shall now examine the response to small perturbations from these launching conditions. As the system is invariant under translations in the $y$-direction, the WGMs are neutrally stable with respect to perturbations of the launching point's $y$-coordinate. As a ray launched in the $\hat{y}$ direction from  $(x_0,y_0,\tau - \delta)$ is identical to a ray launched from $(x_0,y_0- \delta,\tau)$, WGMs are also neutrally stable with respect to perturbations of the launching point's $z$ coordinate. This indicates the existence of a band of WGMs for all values of $x$ between $\frac{1}{4\tau^2}$ and $1$. A ray launched in the $\hat{y}$ direction outside of this band will get trapped onto a wave-attractor parallel to the $(x,z)$ plane. This proves that WGMs are also neutrally stable with respect to perturbations of the launching point's $x$ coordinate.

If instead of perturbing the launching parameters, the perturbation is in the wave frequency, one may see it as a perturbation of the basin itself. Specifically this will augment its effective height $\tau$, after the re-stretching of the vertical coordinate to achieve $\gamma=1$. In that case the ray's trajectory is still a WGM, however, its step size has changed due to the change in $\tau$.

\subsubsection{Angular perturbations}

\begin{figure}
  \centerline{\includegraphics[width=13cm]{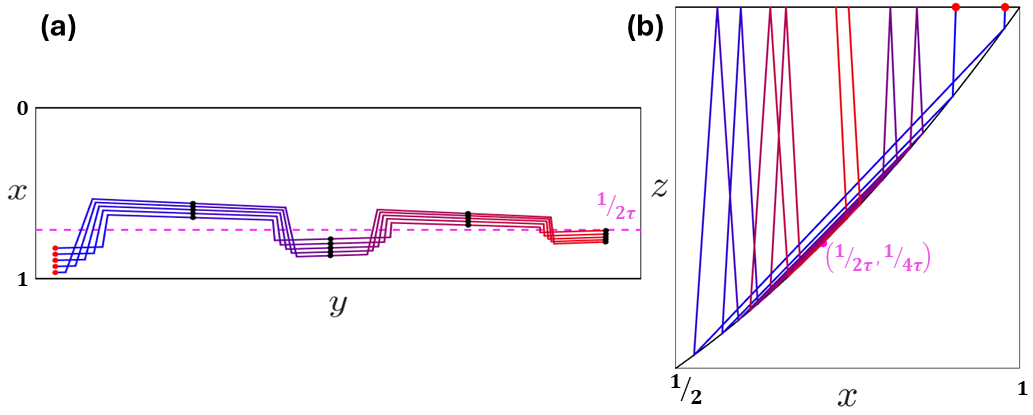}}% Images in 100% size
  \caption{(a) Top view of internal wave beam in a parabolic channel of scaled depth $\tau=0.7$ launched inwards with respect to the WGM. Red points indicate launching locations and black points represent reflections from the channel's surface. The magenta critical line is dashed. (b) Side view of two rays focusing after being launched inwards with respect to the WGM.
  In both figures particle velocities increase, as implied by the focusing of the ray. Color qualitatively highlights diminishing distance between rays in a beam.}
\label{fig:6}
\end{figure}

We now go on to numerically examine rays with a perturbed launching angle $\varphi = \frac{\pi}{2} + \epsilon$ with respect to the $\hat{x}$ direction.

We observe a ray launched `outwards' at the surface, from $x>x_{crit}$, and with $\epsilon < 0$. Upon reflection from the bottom it will refract towards the rim. Upon subsequent reflections, the ray will increase moving into cross-slope direction until it  gets trapped by a wave-attractor. This is unsurprising as wave-attractors in channels are well documented \citep{manders2004three,drijfhout2007impact,pillet2019internal,bratspiess2025nonspecular}. We now observe a ray launched `inwards' from the same location, i.e., having $\epsilon>0$. This time the ray will drift inwards, as depicted in Fig.~\ref{fig:6}, in the direction of the critical slope. As long as $|\epsilon|<\epsilon_{crit}$, determined below, the ray will be trapped at and above the critical slope, on a two-dimensional wave attractor that unlike those known up until now lays parallel to the channels direction instead of perpendicular to it.

 As long as a ray launched from the surface intersects the channel's bottom on the same side of the critical line as its launching point, it is in the critical slope attractor's basin of attraction. In order to find $\epsilon_{crit}$ for a ray launched from $(x_0,1)$ we calculate the critical effective 2D inclination for which the ray hits exactly the critical point $(\frac{1}{2\tau},\frac{1}{4\tau})$.
\begin{equation}
    \gamma_{crit}'=\frac{4\tau-1}{2\tau x_0-1}\Rightarrow\frac{\pi}{2}+\epsilon_{crit}=\arccos{(-\frac{2\tau x_o - 1}{4\tau - 1})}
\end{equation}

We conclude that the border between trapping at the critical-slope attractor or at a classical cross-channel attractors is given by rays launched exactly down-channel, which constitute the WGMs.

When a beam of parallel rays is launched inwards and drifts towards the critical level they move closer to each other and become denser. From conservation of momentum it follows that upon  focusing of the internal wave beam its amplitude increases. From this we can expect increased velocities and energy accumulation around the critical slopes.

At the critical slope itself the ray dynamics breaks down. The asymptotic analysis in \citep{bratspiess2025nonspecular} predicts that at the critical level a ray coming from above will be focused perfectly along the cross-channel $x$-direction, towards the center of the channel. However this trajectory is unavailable due to the channel being convex, rendering the trajectories no longer physically possible. Once the rays approach the vicinity of the critical slope, velocities diverge and the local length scale goes to zero indicating that the original scaling analysis no longer holds and nonlinear and dissipative effects are no longer negligible. These new effects will smooth out the consequences of the dynamics' breakdown, similar to the asymptotic theory in \cite{he2022internal}.

The velocity profile of a ray is not uniform along the WGM cycle. From ray densities in Fig.~\ref{fig:6} we can deduce that particle velocities are lowest in the outer vertical section of the orbit, where rays are most sparse, and highest in the diagonal section where rays are the most dense. In the same manner, the focusing and velocity amplification is not monotonic during the whole focusing depicted in Fig.~\ref{fig:6} but only when comparing the same sections in different periods. Wave group and phase velocities go to zero as rays approach the critical slope

\section{Internal wave rays in trapezoid channels}

\subsection{WGM existence}
The cross section of a trapezoidal channel is defined by three parameters, stretched height $\tau$, base length $a$, and wall slope $s = \tan{\mu}$. By scaling coordinates one can chose, without loss of generality, $\tau=1$. This rescaling leaves the two other parameters as sole definers of channel geometry. To avoid rays getting trapped in corners we focus on channels with $s>1$.

Similar to the WGMs in the previous subsection, we once again examine rays launched down-channel from the rigid lid surface. Rays launched from above the flat base are uninteresting, so we focus on those launched above the sloping wall. As vertical and horizontal walls reflect IW rays specularly, we use the corners in which they intersect as mirrors to achieve the desired periodic orbits. 

A ray launched above the sloping wall parallel to the $y$-direction will be reflected to a horizontal angle related to the effective inclination
\begin{equation}
    \gamma'=\frac{1+s^2}{2s}
\end{equation}
To derive the geometrical condition restricting the existence of WGMs we examine its potential central ray, a ray launched from a corner with the 2D effective inclination from equation (3.6). A WGM exists if this ray hits the sloping wall from below, effectively backtracking its path, as described in the previous paragraph. A simple geometric calculation yields the condition
\begin{equation}
    a\frac{s^2+1}{2s} - \left \lfloor{ a\frac{s^2+1}{2s}}\right \rfloor <\frac{s^2-1}{2s^2}
\end{equation}
that, when satisfied, indicates that a WGM indeed exists. 
\begin{figure}[h]
  \centerline{\includegraphics[width=13.6cm]{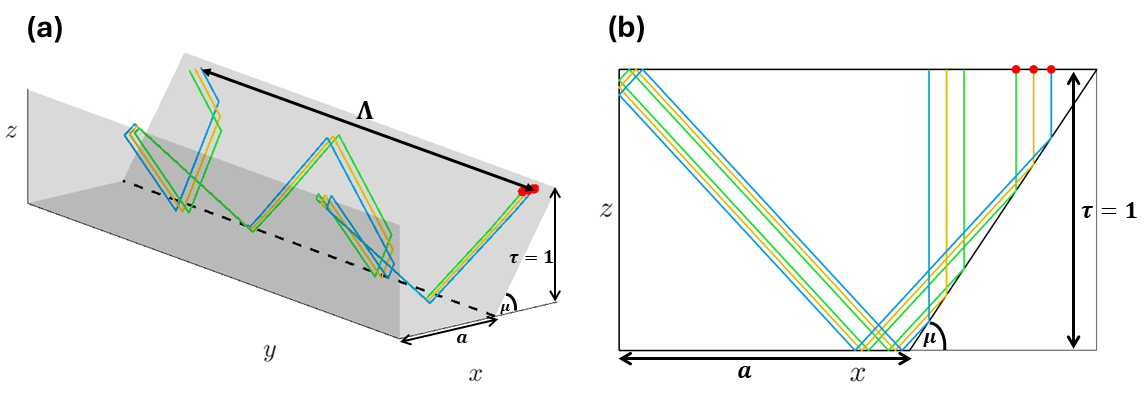}}% Images in 100% size
  \caption{(a) WGMs in a trapezoid channel launched from the red dots parallel to the $y$-direction. All WGMs share $M=1$ (b) Two-dimensional projection of the first half of the channel depicted in (a) onto the $x-z$ plane. Vertical lines correspond to rays propagating parallel to the channel.  }
\label{fig:3}
\end{figure}
As in the previous subsection, all WGMs in a trapezoidal channel share a step size:
\begin{equation}
\Lambda = 1+M\frac{s^2+1}{s^2-1}-as\frac{2}{s^2-1}    
\end{equation}
where $M$ is the number of bottom and surface reflections before the ray  reflects specularly from the corner. Figure \ref{fig:3} shows an example, see also \citep{manders2004three}.

\subsection{Stability analysis}
Similar to the behavior seen in a parabolic channel, WGMs in a trapezoidal channel are neutrally stable with respect to spatial perturbations and unstable with respect to angular perturbations at launch.

Unlike parabolic channels, where horizontal angle instability was demonstrated by numerical simulations, in the trapezoidal channel instability can be understood analytically. A ray launched in the channel having a single inclined wall has only one  slope, $s$, to focus and de-focus from. This means that reflections from above and below are reciprocal in focusing and de-focusing terms. In that case a ray launched at a small angle $\epsilon$ with respect to the along-channel $\hat{y}$ direction preserves its original horizontal angle after a full WGM cycle. This means that even with a small angle the ray will consistently drift away from the periodic WGM, eventually becoming trapped at a wave attractor in a plane orthogonal to the sloping wall. 

This means that unlike parabolic channels where WGMs are at the border between two basins of attraction of two different types of wave attractors, in trapezoid channels, where only a single basin of attraction exists, WGMs can be seen as its boundary having infinite focusing time.

\subsection{Whispering Gallery Beams}
Since both in parabolic as well as trapezoidal channels we can find continuous groups of rays that propagate together while remaining coherent, in both cases we can define \textit{Whispering Gallery Beams}. Similar to a beam of light, these beams can transport energy and information coherently across long distances. The difference between standard beams and WG beams is that the second type's existence depends on the existence of a critical slope or corner  acting as a critical slope. 

\section{Relating observations with ray tracing simulations}

Our analysis of Internal Wave Whispering Gallery Modes lets us intuitively understand several phenomena in submarine canyons and open ocean basins, specifically along-canyon tidal energy fluxes, similar to ones measured in \citep{ASLAM201820}, and energy and turbulence accumulation, similar to what was observed in \citep{Horn_Meincke_1976} and \citep{nash2004internal}.
\subsection{Internal tide energy fluxes in submarine canyons}
Similar to how electromagnetic waves propagate through optical fibers \citep{Senior_Jamro_2009} and whale sounds propagate across the ocean through the SOFAR channel \citep{payne1971orientation,braun2022functional}, so can internal waves propagate along submarine canyons via WGMs, albeit with different mechanisms enabling their propagation. Although our theoretical analysis deals with idealized parabolic channels, the dynamics of real ocean currents in submarine canyons can be approximated near their critical levels. There, channel slopes match ray inclinations and the channel can locally be seen as straight in along-channel and parabolic in cross-channel directions. 

As WGMs enable attenuation-less propagation, one expects that even localized forcing will lead to an energy distribution all along the channel dispersed by said WGMs. Numerical simulations by \cite{drijfhout2007impact} and ocean measurements by \cite{ASLAM201820} further strengthen this hypothesis. 

\begin{figure}
  \centerline{\includegraphics[width=14cm]{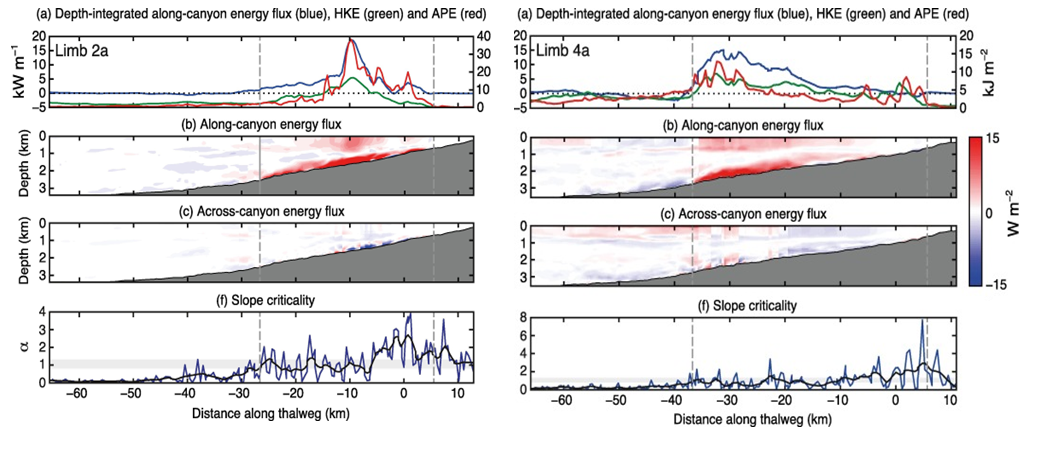}}% Images in 100% size
  \caption{adaptations of fig 6 (left) and fig 7 (right) of \citep{ASLAM201820}. (a) Depth-integrated along-canyon baroclinic
  M2 energy flux (blue), HKE (green), and APE (red) with
  distance along the thalweg. (b) Along-canyon and (c)
  across-canyon baroclinic M2 energy flux with distance
  along the thalweg. Positive along-canyon values are
  toward the head of the canyon limb. Positive acrosscanyon
  values are to the left when looking up canyon. (f) Along-thalweg slope criticality to the
  M2 internal tide (blue) and smoothed using a 5-km
  running mean (black). Near-critical values
  ($0.8 < \alpha < 1.3$; \cite{mcphee2002boundary}) are
  indicated in grey. The dashed grey lines indicate the
  distance over which bottom intensification occurs.}
\label{fig:5}
\end{figure}

\cite{ASLAM201820} measured internal tides in the Bay of Biscay, and investigated their along and across channel energy fluxes. They observed that along-canyon energy fluxes coincide with channel wall slopes being near-criticality with respect to the semidiurnal tidal frequency, as is now predicted, which could not be explained by across-isobath wave attractors. These measurements, presented in Fig.~\ref{fig:5}, also support our claim that due to lack of along-channel flux at locations where channel geometry does not enable WGMs, energy can propagate along the channel. Lack of across-canyon energy fluxes outside of the critical vicinity of the channel support our theory in the sense that no other mechanism is needed in order to explain these results.

\subsection{Energy accumulation near critical slopes}
\begin{figure}
  \centerline{\includegraphics[width=8cm]{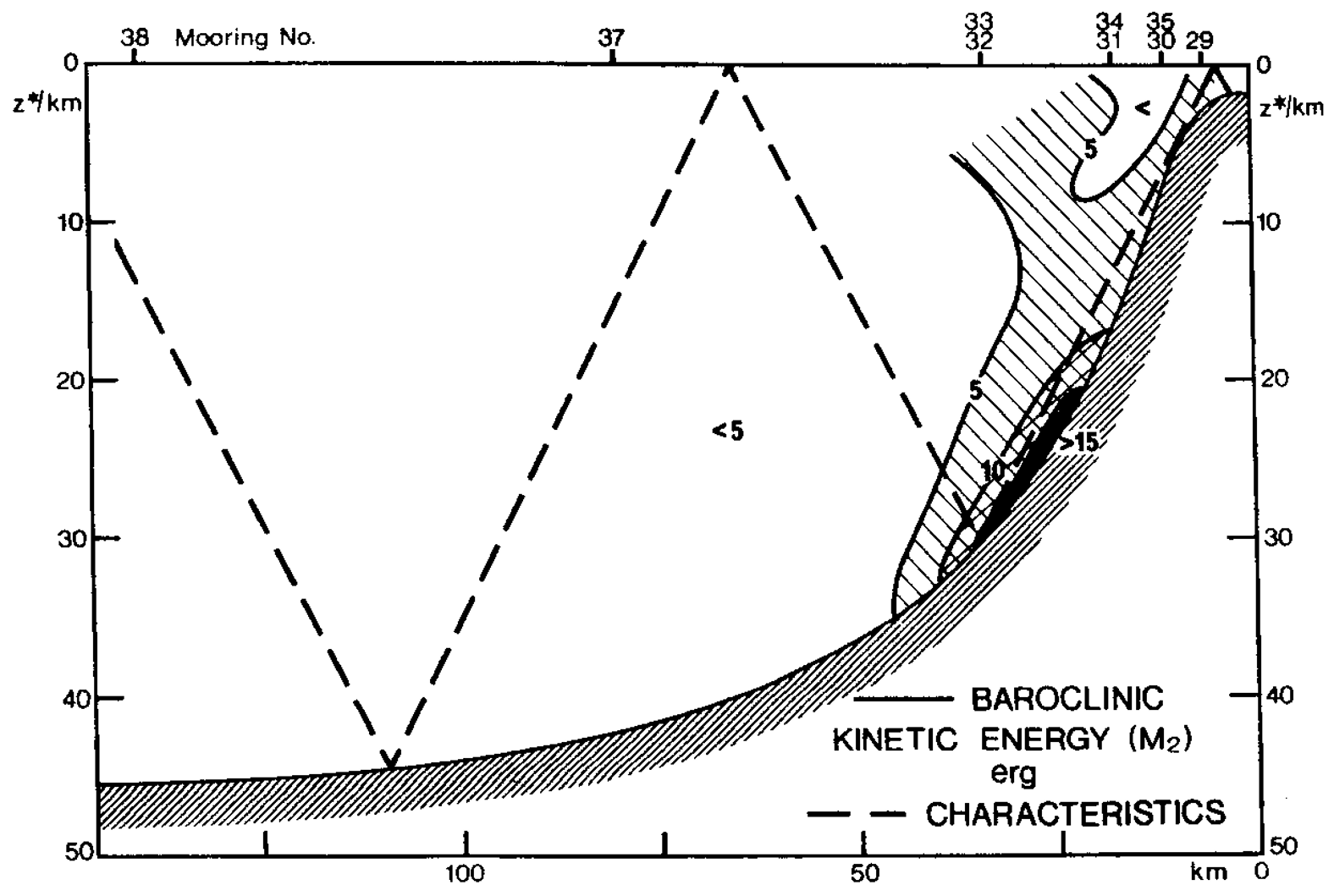}}% Images in 100% size
  \caption{Characteristics and distributions  of baroclinic tidal energy on a continental slope section. According to the local stratification rate, WKB-stretching of the vertical has been applied. This transforms ray paths into characteristics of constant slope. Taken from \citep{Horn_Meincke_1976}.}
\label{fig:7}
\end{figure}
The new mechanism of focusing of internal waves not towards the well known wave attractors but rather towards the critical slope can explain the accumulation of energy, turbulence, and non-linear interactions near critical slopes. The diverging ray density near the critical point seen in Fig.~\ref{fig:6}b, indicating diverging velocities and pressure, is reminiscent of the accumulations observed in nature \citep{Horn_Meincke_1976, nash2004internal}, shown in Fig.~\ref{fig:7}, and in numerical simulations \citep{drijfhout2007impact}. Despite the fact that critical slope energy accumulations may be explainable by across-isobath wave attractors, we find it less likely in the settings of continental slopes such as in \citep{Horn_Meincke_1976} as the lack of horizontal closure allows only along-isobath critical slope wave attractors. 

\section{Conclusion}
To conclude, via direct geometric construction we have analytically proved the conjectured existence of internal wave \textit{Whispering Gallery Modes} (WGMs), i.e., periodic orbits in the three-dimensional internal wave-ray billiard that do not get absorbed by any wave-attractor, in parabolic and trapezoidal channels. Utilizing similar techniques we have done the same in a rotationally invariant truncated cone basin. 

In channels, we have shown that WGMs are neutrally stable with respect to perturbations in launching location. From this we deduce the existence of \textit{Whispering Gallery Beams}, continuous groups of rays propagating coherently through the channels. These whispering gallery beams can be used to transport energy and information similarly to electromagnetic or acoustic waves in different waveguides.

By examining perturbations in launching angles we have discovered a new kind of wave attractor in parabolic channel. The new wave attractor discovered lays along the critical line and above it, counter to the channel attractors known until now which lay perpendicular to the channel's direction. It is worth noting that although the channel basins analyzed are invariant along one of their axes, this phenomenon is purely three-dimensional and cannot be reproduced in two-dimensional basins shaped as cross-sections of the full three-dimensional channels.

We suggest that these whispering gallery beams and critical-slope attractors may explain several phenomena observed in nature. Specifically, whispering gallery beams may account for along-channel energy fluxes observed by \cite{ASLAM201820} and critical-slope attractors as a mechanism explaining accumulations of energy and non-linearities in the vicinity of critical slopes.

One of the future directions the research may take is answering the unresolved question regarding the fate of non-converging rays in a paraboloidal basin introduced by \cite{maas2005wave}. A potential solution to this question may be that some rays approach a circular critical-slope attractor, bound by circular WGMs, similar to those in the parabolic channel. However, further research is required.

Another potential future direction is a revision of previous works in which three-dimensional basins are approximated by two-dimensional ones. Re-examinations may reveal phenomena unexpectedly hidden by the simplification, even though intuitively the two and three-dimensional systems seem equivalent. A specific basin that may be worthwhile to re-investigate is the spherical shell, in which an examination of meridional cross-sections and across-isobath wave attractors, might miss WGMs and critical slope attractors hidden by the two-dimensional projection.

\backsection[Acknowledgments]{N. Bratspiess wishes to thank Ilias Sibgatullin for insightful discussions which helped re-imagining the analysis.}

\appendix

\section{}\label{appA}
Now that we have concluded the analysis of straight channels, basins invariant with respect to a single Cartesian coordinate, we move on to basins invariant with respect to the angular coordinate. Rotationally invariant basins are common in the literature on internal waves \citep{maas2005wave, rabitti2013meridional,rabitti2014inertial, sibgatullin2017direct, pacary2023observation}, and are, among other reasons, relevant in the context of stellar dynamics in astrophysics \citep{RIEUTORD_VALDETTARO_2010, ogilvie2013tides,LEBARS201048}.

\subsection{2D projection}
In the polar coordinate system, $(\rho,\theta,z)$, rotation bodies such as cones, spheres, paraboloids, and others can be represented as single variable function $\rho(z)$ or $z(\rho)$. Intuitively, this persuades us to look for the effective 2D billiard of the 3D systems, an example of which is depicted in Fig.~\ref{fig:4}. We shall do so by examining a differential step of a ray propagating at inclination $\pm 1$. In the horizontal plane a straight path satisfies

\begin{equation}
    \rho=\frac{\alpha}{\cos{(\theta-\theta_0)}}
\end{equation}
Where $\alpha$ and $\theta_0$ are determined by the ray's launching location and direction. The horizontal differential step satisfies

\begin{equation}
    \vec{dl} =(d\rho,\rho d\theta) \Rightarrow \gamma^{- 1} dz=|\vec{dl}|=\sqrt{1+\rho^2(\frac{d\theta}{d\rho})^2}d\rho=\frac{\rho}{\sqrt{\rho^2-\alpha^2}}d\rho.
\end{equation}
From this we calculate the curve which the ray follows in $(\rho,z)$-space:

\begin{figure}
  \centerline{\includegraphics[width=12cm]{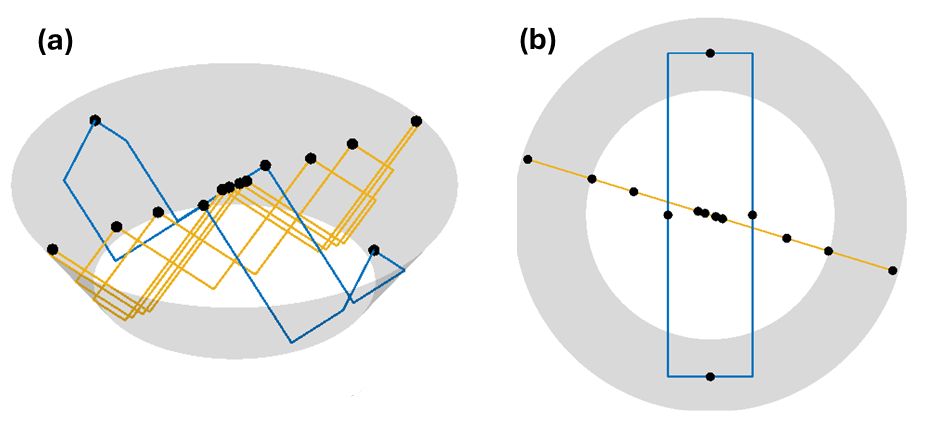}}% Images in 100% size
  \caption{(a) Side view of a Wave-attractor (yellow) and a WGM (blue) in a frustum basin. surface reflections indicated by black dots. (b) Top view of (a)}
\label{fig:4}
\end{figure}

\begin{equation}
    \rho^2=\alpha^2+[{\gamma^{- 1}(z-z_0)+\sqrt{\rho_0^2-\alpha^2}}]^2.
\end{equation}
The intersection of this hyperbola with the boundary is the next collision point, and the new angle $\theta$ can be calculated from equation (4.1). Although this projection in much less elegant due to the projected characteristics being hyperbolas and not straight lines, it is still a simplification of sorts which may be found useful in the future. 

\subsection{Beyond the 2D projection}
Due to the awkwardness of the 2D projection explored in the previous subsection it is sometimes more convenient to approach the 3D billiard directly. An example for one such case is the frustum, a truncated cone which can be defined by floor radius $R_{in}$, ceiling radius $R_{out}$, and stretched height $\tau$. $s = \frac{\tau}{R_{out}-R_{in}}$ is once again the wall's slope. In the frustum  one can find periodic or quasi-periodic WGMs by imposing a condition on the sum of incoming and outgoing angles, and then finding the reflection radius on the frustum that results in the desired trajectory. One such WGM is depicted in Fig.~\ref{fig:4}.

Following the same approach, in a paraboloidal basin this results in a bowtie-WGM similar to the ones in \citep{rabitti2014inertial}. This approach can also be used in the case of basins with discrete symmetries, such as the truncated elliptic cone in \citep{favier2024inertial} or the 3D stadium in \citep{bratspiess2025nonspecular}. In these cases no simple projection to two dimensions can exist, however the existing symmetries may still enable us to find WGMs.

\bibliographystyle{jfm}
\bibliography{jfm}

\end{document}